# Giant thermal Hall conductivity from neutral excitations in the pseudogap phase of cuprates

G. Grissonnanche[1], A. Legros[1,2], S. Badoux[1], E. Lefrançois[1], V. Zatko[1], M. Lizaire[1], F. Laliberté[1], A. Gourgout[1], J.-S. Zhou[3], S. Pyon[4], T. Takayama[4], H. Takagi[4], S. Ono[5], N. Doiron-Leyraud[1], and L. Taillefer[1,6]

*1 Institut quantique, Département de physique & RQMP, Université de Sherbrooke, Sherbrooke, Québec J1K 2R1, Canada*

*2 SPEC, CEA, CNRS-UMR3680, Université Paris-Saclay, Gif-sur-Yvette Cedex 91191, France*

*3 Materials Science and Engineering Program, Mechanical Engineering, University of Texas at Austin, Austin, Texas 78712, USA*

*4 Department of Advanced Materials, University of Tokyo, Kashiwa 277-8561, Japan*

*5 Central Research Institute of Electric Power Industry, Kanagawa 240-0196, Japan*

*6 Canadian Institute for Advanced Research, Toronto, Ontario M5G 1Z8, Canada*

**The nature of the pseudogap phase of cuprates remains a major puzzle. Although there are indications that this phase breaks various symmetries, there is no consensus on its fundamental nature[1]. Although Fermi-surface[2], transport[3] and thermodynamic[4] signatures of the pseudogap phase are reminiscent of a transition into a phase with antiferromagnetic order[5,6], there is no evidence for an associated long-range magnetic order. Here we report measurements of the thermal Hall conductivity $\kappa_{xy}$ in the normal state of four different cuprates ($La_{1.6-x}Nd_{0.4}Sr_xCuO_4$, $La_{1.8-x}Eu_{0.2}Sr_xCuO_4$, $La_{2-x}Sr_xCuO_4$, and $Bi_2Sr_{2-x}La_xCuO_{6+\delta}$) and show that a large negative $\kappa_{xy}$ signal is a property of the pseudogap phase, appearing with the onset of that phase at the critical doping $p^*$. Since it is not due to charge carriers – as it persists when the material becomes an insulator, at low doping – or magnons – as it exists in the absence of magnetic order – or phonons – since skew scattering**

**is very weak, we attribute this $\kappa_{xy}$ signal to exotic neutral excitations, presumably with spin chirality[7]. The thermal Hall conductivity in the pseudogap phase of cuprates is reminiscent of that found in insulators with spin-liquid states[8,9,10]. In the Mott insulator $La_2CuO_4$, it attains the highest known magnitude of any insulator[11].**

Among the different families of unconventional superconductors, magnetism and superconductivity are often in close proximity[12]. A notable exception is hole-doped cuprates, where instead superconductivity mostly coexists with the pseudogap phase, an enigmatic state of matter whose nature remains unclear[1]. The doping $p^*$ for the onset of the pseudogap phase bears the hallmark of an antiferromagnetic quantum critical point[13], with a sharp drop in the carrier density $n$, from $n \sim 1 + p$ above $p^*$ to $n \sim p$ below[3,14], a $T$-linear resistivity[14], and a $\log(1/T)$ specific heat[4]. Yet, there is no evidence for long-range magnetic order appearing at $p^*$. However, numerical solutions to the Hubbard model have shown that a pseudogap phase can arise from short-range antiferromagnetic correlations[15]. It has been argued that an exotic state with topological order can account for such a pseudogap and for the drop in carrier density without breaking translational symmetry[16], but the low-energy excitations of such a state have yet to be detected.

In recent years, the thermal Hall effect has emerged as a powerful probe of magnetic texture and topological excitations in insulators. On the theory side, a non-zero thermal Hall conductivity $\kappa_{xy}$ was shown to arise even without long-range magnetic order, either from the spin chirality of a paramagnetic state[7] or from fractionalized (topological) excitations in a spin liquid[17]. On the experimental side, a sizable $\kappa_{xy}$ has been measured in insulators without magnetic order, such as the spin-ice system $Tb_2Ti_2O_7$ (ref. 18) and the spin-liquid systems $RuCl_3$ (ref. 8), volborthite[9] and Ca kapellasite[10].

In cuprates, studies of $\kappa_{xy}$ have so far been limited to the superconducting state[19,20,21], except for the case of YBa$_2$Cu$_3$O$_y$ (YBCO) at $p = 0.11$, where $\kappa_{xy}$ was measured in the field-induced normal state[22], which has charge-density-wave order[13]. See Methods for a discussion of this particular case.

Here, we investigate the thermal Hall response of the pseudogap phase via measurements of $\kappa_{xy}$ in four different cuprate materials – La$_{2-x}$Sr$_x$CuO$_4$ (LSCO), La$_{1.6-x}$Nd$_{0.4}$Sr$_x$CuO$_4$ (Nd-LSCO), La$_{1.8-x}$Eu$_{0.2}$Sr$_x$CuO$_4$ (Eu-LSCO) and Bi$_2$Sr$_{2-x}$La$_x$CuO$_{6+\delta}$ (Bi2201) – across a wide doping range, from the overdoped metal at $p = 0.24$ down to the Mott insulator at $p = 0$ (Fig. 1a). The $\kappa_{xy}$ data reported here are all in the normal state, with superconductivity suppressed by application of a magnetic field normal to the CuO$_2$ planes.

In Nd-LSCO and Eu-LSCO, the critical doping for the onset of the pseudogap phase is at $p^* = 0.23$ (refs. 4, 13, 14) (Fig. 1a). In Fig. 2a, we plot $\kappa_{xy} / T$ vs $T$ for Nd-LSCO at $p = 0.24$: $\kappa_{xy}$ is positive and $\kappa_{xy} / T$ increases monotonically with decreasing $T$, tracking closely the electrical Hall conductivity $\sigma_{xy}$ measured on the same sample, satisfying the Wiedemann-Franz law as $T \to 0$, namely $\kappa_{xy} / T = L_0 \sigma_{xy}$, where $L_0 = (\pi^2/3)(k_B/e)^2$. The large positive value of $\sigma_{xy}$ is dictated by the large Fermi surface at $p > p^*$ and its Hall number $n_H \sim 1 + p$ (ref. 14). Clearly, at $p = 0.24$, $\kappa_{xy}$ is due to charge carriers.

We now turn to dopings immediately below the pseudogap critical point. In Fig. 2b, we plot $\kappa_{xy} / T$ vs $T$ for Nd-LSCO at $p = 0.20$. We see a qualitatively different behavior, with $\kappa_{xy}$ becoming negative at low $T$. As seen in Fig. 3a, this qualitative change occurs immediately below $p^*$. In Eu-LSCO, the very same change occurs across $p^*$ (Fig. 3b), from positive $\kappa_{xy}$ above $p^*$ ($p = 0.24$) to negative $\kappa_{xy}$ (at low $T$) below $p^*$ ($p = 0.21$), with essentially identical data to Nd-LSCO at $p = 0.24$ and $p = 0.21$. The negative $\kappa_{xy}$ is therefore a property of the pseudogap phase.

We also measured $\kappa_{xy}$ in Bi2201, a cuprate with a different crystal structure to that of Nd-LSCO and Eu-LSCO, on an overdoped sample of La content $x = 0.2$, with $p$ slightly below $p^*$ (ref. 23). In Fig. 2d, we see that $\kappa_{xy}(T)$ in Bi2201 displays a remarkably similar behavior to that of Nd-LSCO and Eu-LSCO at $p < p^*$. A negative thermal Hall conductivity $\kappa_{xy}$ at low temperature is therefore a generic property of the pseudogap phase, independent of material. Note that the electrical Hall conductivity $\sigma_{xy}$ measured on the same samples remains positive down to $T \rightarrow 0$ (Figs. 2b, 2d).

We now move to much lower doping. In Fig. 1b, we see that $\kappa_{xy} / T$ is still negative at low temperature in Eu-LSCO at $p = 0.08$ and in LSCO at $p = 0.06$, where in both cases $\sigma_{xy}$ is positive and completely negligible (Figs. 2e, 2f), because the samples are almost electrically insulating at low temperature. This shows that the negative $\kappa_{xy}$ signal of the pseudogap phase is not due to charge carriers.

Magnons can also be excluded as the source of the negative $\kappa_{xy}$. In the phase diagram of Fig. 1a, we delineate in gray the regions where static magnetism is detected by μSR, whether as incommensurate correlations below $T_m$ or as commensurate Néel order below $T_N$. We see that in all three materials – Nd-LSCO at $p = 0.20$, Eu-LSCO at $p = 0.08$ and LSCO at $p = 0.06$ – the negative $\kappa_{xy}$ signal is present well above $T_m$ (Fig. 1), where there is no static magnetism. Moreover, the $\kappa_{xy}(T)$ curve for $La_2CuO_4$ (Fig. 1b), i.e. undoped LSCO with $p = 0$, where there is long-range antiferromagnetic order below ~ 300 K (Fig. 1a), is very similar to the curve for LSCO at $p = 0.06$ (Fig. 1b), where there is no magnetic order above $T \sim 5$ K (Fig. 1a). (See Methods for further discussion of magnons.) We conclude that magnetic order is not responsible for the negative $\kappa_{xy}$ signal seen in cuprates at all dopings below $p^*$, and magnons are ruled out as the relevant excitations.

Phonons can generate a non-zero $\kappa_{xy}$ signal if they are subject to skew scattering by spins[24]. Spin scattering will also show up in the longitudinal thermal conductivity $\kappa_{xx}$, which is dominated by phonons, in two ways: 1) it reduces the magnitude of $\kappa_{xx}$ relative to a non-magnetic analog material; 2) it produces a field dependence in $\kappa_{xx}$, whose strength is measured by the ratio $[\kappa_{xx}(H) - \kappa_{xx}(0)] / \kappa_{xx}(0)$. Let us compare LSCO ($p = 0.06$) to two insulators with strong spin scattering of phonons (and no magnetic order): $Tb_2Ti_2O_7$ (ref. 18) and $Ba_3CuSb_2O_9$ (ref. 25). In the latter two materials, $\kappa_{xx} / T \sim 25$ mW / K$^2$ m at $T = 15$ K, compared to $\kappa_{xx} / T \sim 300$ mW / K$^2$ m in LSCO (Extended Data Fig. 1d), a massive reduction due to strong spin scattering (see Methods and Extended Data Fig. 2a). The field dependence of $\kappa_{xx}$ is correspondingly much weaker in LSCO (Extended Data Fig. 2): $[\kappa_{xx}(H) - \kappa_{xx}(0)] / \kappa_{xx}(0) \sim 0.4$ % in LSCO (at $T = 15$ K, $H = 15$ T), compared to 32 % in $Tb_2Ti_2O_7$ (at $T = 15$ K, $H = 8$ T) and 5 % in $Ba_3CuSb_2O_9$ (at $T = 5$ K, $H = 15$ T) (see Table 1). One would therefore expect a much smaller $\kappa_{xy}$ signal from phonons in LSCO, but in fact $\kappa_{xy}$ in LSCO is much larger. In absolute terms, $|\kappa_{xy} / T| = 2$ mW / K$^2$ m in LSCO (Fig. 1b), compared to $\sim 0.08$ mW / K$^2$ m in $Tb_2Ti_2O_7$ and $\sim 0.002$ mW / K$^2$ m in $Ba_3CuSb_2O_9$ – so 20 to 1000 times larger. In relative terms, $|\kappa_{xy} / [\kappa_{xx}(H) - \kappa_{xx}(0)]| \sim 1$ in LSCO, compared to $\sim 0.01$ in $Tb_2Ti_2O_7$ and $\sim 0.002$ in $Ba_3CuSb_2O_9$ (Table 1) – so 100 to 300 times larger. We conclude that phonons are so weakly scattered by spins that they cannot cause the huge $\kappa_{xy}$ signal in LSCO at $p = 0.06$ (or Eu-LSCO at $p = 0.08$ or $La_2CuO_4$).

Moreover, the phonon conductivity in Nd-LSCO shows no indication that strong spin scattering sets in abruptly below $p^*$, to act as a phonon mechanism for the negative $\kappa_{xy}$ signal that suddenly appears below $p^*$. Indeed, $\kappa_{xx}$ does not decrease below $p^*$, on the contrary, it increases (Extended Data Fig. 3), most likely because electron-phonon scattering decreases as the charge carrier density drops. We conclude that phonons are not responsible for the large negative $\kappa_{xy}$ signal of cuprates that appears suddenly below $p^*$. (See Methods for further discussion.)

The $\kappa_{xy}$ signal in the Mott insulator $La_2CuO_4$ (and in LSCO at $p = 0.06$) is the largest ever seen so far in any insulator. Only multiferroic materials like ferrimagnetic $(Fe,Zn)_2Mo_3O_8$ have comparable $\kappa_{xy}$ values[11] (Fig. 4b), thanks to their exceptionally strong lattice-spin coupling. That the underlying mechanism is completely different in cucprates and multiferroics can be seen by the field dependence of $\kappa_{xx}$, a direct measure of the lattice-spin coupling, which is ~ 100 times larger in $(Fe,Zn)_2Mo_3O_8$ (Fig. 4a).

The large negative $\kappa_{xy}$ reported here for cuprates is not due to electrons, magnons or phonons. It must come from as yet unidentified neutral excitations. Identifying these excitations will shed new light on the nature of the pseudogap phase. It is instructive to compare cuprates with insulators that are believed to host spin-liquid states. The largest $\kappa_{xy}$ signal so far in such materials was detected in $RuCl_3$ (Fig. 4b). In this 2D material, spins on a honeycomb lattice are frustrated and only order (antiferromagnetically) below $T_N = 7$ K. Above $T_N$, the paramagnetic state is thought to be a spin liquid state described by the Kitaev model[17]. In Fig. 4c, we reproduce the data of Hentrich et al.[26] for $\kappa_{xy} / T$ vs $T$ in $RuCl_3$. Above 100 K, $\kappa_{xy} / T$ is vanishingly small. Below 100 K, $\kappa_{xy} / T$ grows gradually with decreasing $T$ down to 20 K or so (and then drops rapidly as $T_N$ is approached). In the regime between 20 K and 100 K, $\kappa_{xy} / T$ is well described by calculations for the Kitaev model[17], implying that the $\kappa_{xy}$ signal in $RuCl_3$ comes from itinerant Majorana fermions – exotic neutral excitations of topological character. This interpretation is supported by the observation[27] of a predicted[17] quantization of the thermal Hall conductivity (at low $T$ when AF order is suppressed by applying a field in the 2D planes). Other spin-liquid candidates, like volborthite[9] and Ca kapellasite[10], exhibit qualitatively similar $\kappa_{xy}(T)$ (Fig. 4d), suggesting that the gradual growth below ~ 100 K is a general behavior.

In Figs. 4c and 4d, we compare our data on LSCO $p = 0.06$ to the data on RuCl$_3$ and Ca kapellasite, respectively. There is a tantalizing similarity in the gradual growth below 100 K or so, but there are some differences. First, whereas $\kappa_{xy}$ is positive in these two spin-liquid candidates, it is negative in cuprates. (This may reflect the particular topological character of the different states.) Secondly, the signal in LSCO is ~ 10 to 25 times larger (Fig. 4). Finally, in LSCO, $\kappa_{xy} / T$ continues to grow down to the lowest measured temperature – but it may well drop below ~ 5-10 K.

In summary, the thermal Hall effect in cuprates reveals a hitherto unknown facet of the enigmatic pseudogap phase, reminiscent of a spin liquid. It points to a state with chirality[7]. It will be interesting to see whether models of spin-charge separation[28], topological order[16] or current loops[29], for example, may be consistent with the giant $\kappa_{xy}$ signal that appears below $p^*$.

**Acknowledgements.** L.T. acknowledges support from the Canadian Institute for Advanced Research (CIFAR) and funding from the Natural Sciences and Engineering Research Council of Canada (NSERC), the Fonds de recherche du Québec - Nature et Technologies (FRQNT), the Canada Foundation for Innovation (CFI), and a Canada Research Chair. This research was undertaken thanks in part to funding from the Canada First Research Excellence Fund. Part of this work was funded by the Gordon and Betty Moore Foundation's EPiQS Initiative (Grant GBMF5306 to L.T.). J.-S.Z was supported by NSF MRSEC DMR-1720595 in the US.

**Author information.** The authors declare no competing financial interest. Correspondence and requests for materials should be addressed to G.G. (gael.grissonnanche@usherbrooke.ca) or L.T. (louis.taillefer@usherbrooke.ca).

## MAIN FIGURE CAPTIONS

**Fig. 1 | Phase diagram and thermal Hall conductivity of cuprates.**

**a)** Temperature-doping phase diagram of Nd-LSCO, Eu-LSCO and LSCO, showing the antiferromagnetic phase below the Néel temperature $T_N$ and the pseudogap phase below $T^*$ (ref. 30), which ends at the critical doping $p^* = 0.23$ for both Nd-LSCO (refs. 4, 13, 14) and Eu-LSCO (ref. 4). For LSCO, $p^* = 0.18$ (refs. 13, 30). Short-range incommensurate spin order occurs below $T_m$, as measured by μSR on Nd-LSCO (squares, ref. 31), Eu-LSCO (circles, ref. 32) and LSCO (triangles, ref. 33). The colored vertical strips indicate the temperature range where the thermal Hall conductivity $\kappa_{xy} / T$ at the corresponding doping decreases towards negative values at low temperature (see panel b). **b)** Thermal Hall conductivity $\kappa_{xy} / T$ versus temperature in a field $H = 15$ T, for four materials and dopings as indicated, color-coded with the vertical strips in panel a.

**Fig. 2 | Thermal and electrical Hall conductivities of four cuprates.**

Thermal Hall conductivity $\kappa_{xy}$, plotted as $\kappa_{xy} / T$ (red), and electrical Hall conductivity $\sigma_{xy}$, expressed as $L_0 \sigma_{xy}$ (blue), where $L_0 = (\pi^2/3)(k_B/e)^2$, as a function of temperature in: **a, b)** Nd-LSCO, **d)** Bi2201, **e)** Eu-LSCO, and **f)** LSCO, at dopings $p$ and fields $H$ as indicated. (For Nd-LSCO $p = 0.20$, $\sigma_{xy}$ was measured at $H = 33$ T (ref. 14).) In Nd-LSCO at $p = 0.24$, $\kappa_{xy} / T$ and $L_0 \sigma_{xy}$ are both positive at all temperatures and they track each other, satisfying the Wiedemann-Franz law in the $T = 0$ limit. By contrast, for $p < p^*$ in all four materials, $\kappa_{xy} / T$ falls to large and negative values at low temperature, whereas $L_0 \sigma_{xy}$ remains positive. **c)** Sketch of the thermal Hall measurement (see Methods).

**Fig. 3 | Thermal Hall conductivity across the pseudogap critical point $p^*$.**

Thermal Hall conductivity $\kappa_{xy} / T$ for **a)** Nd-LSCO in $H = 18$ T and **b)** Eu-LSCO in $H = 15$ T, at dopings as indicated, on both sides of the pseudogap critical point $p^* = 0.23$. In both materials, $\kappa_{xy}$ becomes negative at low temperature when $p < p^*$.

**Fig. 4 | Comparison with other insulators, including spin liquid candidates.**

**a, b)** Maximal absolute value of $\kappa_{xy}$ in various insulators, including the multiferroic ferrimagnet $(Fe,Zn)_2Mo_3O_8$ (black diamond; ref. 11) – the previous record holder for the largest $|\kappa_{xy}|$ of any insulator – and the spin-liquid insulator $RuCl_3$ (green squares; refs. 8, 26) – the previous record holder for the largest $|\kappa_{xy}|$ of any insulator without magnetic order. **a)** Maximal $|\kappa_{xy}|$ as a function of the corresponding value of $[\kappa_{xx}(H) - \kappa_{xx}(0)] / \kappa_{xx}(0)$. **b)** Maximal $|\kappa_{xy}|$ as a function of the corresponding $\kappa_{xx}$ value, on a log-log plot. The values for all materials are listed in Table 1. We see that $La_2CuO_4$ and LSCO at $p = 0.06$ have the largest known value of all insulators. **c)** Thermal Hall conductivity $\kappa_{xy} / T$ versus temperature for LSCO at $p = 0.06$ in $H = 15$ T (red) and $RuCl_3$ in $H = 16$ T (blue, ×10; from ref. 26). In $RuCl_3$, the gradual growth of $\kappa_{xy} / T$ upon cooling below $T \sim 100$ K is attributed to Majorana fermions, the topological excitations of the Kitaev spin liquid[8,10,17]. Below $T \sim 20$ K, $\kappa_{xy} / T$ drops upon approaching the antiferromagnetic (AF) phase (grey). **d)** Same as in panel c), for the spin-liquid insulator Ca kapellasite (green, ×25; from ref. 10). Although much larger and negative, the $\kappa_{xy}$ signal in LSCO also comes from neutral excitations in a phase without magnetic order. These comparisons point to a spin-liquid character of the pseudogap phase in cuprates.

| Material | $\kappa_{xy}$ | $\kappa_{xx}$ | $\mid \Delta\kappa_{xx} \mid$ | $\mid \Delta\kappa_{xx}/\kappa_{xx} \mid$ | $T$ | $H$ | Reference |
|---|---|---|---|---|---|---|---|
| | mW / K m | W / K m | W / K m | | K | T | |
| $La_2CuO_4$ | − 38.6 | 12.4 | ∼ 0.06 | ∼ 0.005 | 20 | 15 | this work |
| LSCO | − 30.0 | 5.1 | ∼ 0.02 | ∼ 0.004 | 15 | 15 | this work |
| Eu-LSCO | − 13.2 | 4.5 | ∼ 0.015 | ∼ 0.003 | 15 | 15 | this work |
| $(Fe,Zn)_2Mo_3O_8$ | 24 | 10 | 3.2 | 0.32 | 30 | 9 | 11 |
| $Fe_2Mo_3O_8$ | 24 | 9 | 5 | 0.55 | 45 | 14 | 11 |
| $RuCl_3$ | 8 | 15.5 | 0.62 | 0.04 | 20 | 15 | 8 |
| $RuCl_3$ | 3.5 | 8 | 0.45 | 0.055 | 35 | 16 | 26 |
| $Tb_2Ti_2O_7$ | 1.2 | 0.37 | 0.12 | 0.32 | 15.5 | 8 | 18 |
| Ca kapellasite | 1.1 | 0.2 | --- | --- | 16 | 15 | 10 |
| $Lu_2V_2O_7$ | 1.0 | 0.75 | --- | --- | 50 | 9 | 34 |
| $Ba_3CuSb_2O_9$ | 0.008 | 0.07 | 0.0035 | 0.05 | 5 | 15 | 25 |

**Table 1 | Thermal Hall conductivity in various insulators.**

Maximal value of the thermal Hall conductivity $\kappa_{xy}$ in various insulators, compared to our three cuprates ($La_2CuO_4$, LSCO $p$ = 0.06, Eu-LSCO $p$ = 0.08), measured at the temperature $T$ and field $H$ as indicated: the ferromagnet $Lu_2V_2O_7$ (ref. 34); the multiferroic ferrimagnets $Fe_2Mo_3O_8$ and $(Fe_{0.875}Zn_{0.125})_2Mo_3O_8$ (ref. 11); the spin-ice material $Tb_2Ti_2O_7$ (ref. 18); and the spin-liquid candidates $RuCl_3$ (refs. 8,26), Ca kapellasite[10] and $Ba_3CuSb_2O_9$ (ref. 25). We also list the thermal conductivity $\kappa_{xx}$ measured at the same temperature, in zero field. The change induced in $\kappa_{xx}$ by the field, $\Delta\kappa_{xx} = \kappa_{xx}(H) - \kappa_{xx}(0)$, is given in absolute and relative terms.

# METHODS

SAMPLES

**Nd-LSCO**. Single crystals of $La_{2-y-x}Nd_ySr_xCuO_4$ (Nd-LSCO) were grown at the University of Texas at Austin using a travelling-float-zone technique, with a Nd content $y = 0.4$ and nominal Sr concentrations $x = 0.20, 0.21, 0.22, 0.23$, and $0.25$. The hole concentration $p$ is given by $p = x$, with an error bar ± 0.003, except for the $x = 0.25$ sample, for which the doping is $p = 0.24 ± 0.005$ (for more details, see ref. 14). The value of $T_c$, defined as the point of zero resistance, is: $T_c = 15.5, 15, 14.5, 12$ and $11$ K for samples with $x = 0.20, 0.21, 0.22, 0.23$ and $0.24$, respectively. The pseudogap critical point in Nd-LSCO is at $p^* = 0.23$ (ref. 14).

**Eu-LSCO**. Single crystals of $La_{2-y-x}Eu_ySr_xCuO_4$ (Eu-LSCO) were grown at the University of Tokyo using a travelling-float-zone technique, with a Eu content $y = 0.2$ and nominal Sr concentrations $x = 0.08, 0.21$, and $0.24$. The hole concentration $p$ is given by $p = x$, with an error bar of ± 0.005. The value of $T_c$, defined as the point of zero resistance, is: $T_c = 3, 14$ and $9$ K for samples with $x = 0.08, 0.21$ and $0.24$, respectively. The pseudogap critical point in Eu-LSCO is at $p^* = 0.23$ (ref. 4).

**LSCO**. Single crystals of $La_{2-x}Sr_xCuO_4$ (LSCO) were grown at the University of Tokyo using a travelling-float-zone technique, with nominal Sr concentrations $x = 0.0$ (*i.e.* $La_2CuO_4$) and $0.06$. The hole concentration $p$ is given by $p = x$, with an error bar of ± 0.005. The value of $T_c$, defined as the point of zero resistance, is: $T_c = 0$ and $5$ K for samples with $x = 0.0$ and $0.06$, respectively. The pseudogap critical point in LSCO is at $p^* \sim 0.18$ (ref. 30).

**Bi2201**. Our single crystal of $Bi_2Sr_{2-x}La_xCuO_{6+\delta}$ (Bi2201) was grown at CRIEPI in Kanagawa using a travelling-float-zone technique[35], with La content $x = 0.2$. The value of $T_c$, defined as the onset of the drop in magnetization, is: $T_c = 18$ K. Given its $x$ and $T_c$ values, the doping of this overdoped sample is such that $p < p^*$ (ref. 23).

TRANSPORT MEASUREMENTS

Our comparative study of heat and charge transport was performed by measuring the thermal Hall conductivity $\kappa_{xy}$ and the electrical Hall conductivity $\sigma_{xy}$ on the same sample, using the same contacts made of silver epoxy H20E annealed at high temperature in oxygen.

**Thermal measurements.** A constant heat current $Q$ was sent in the basal plane of the single crystal (along $x$), generating a longitudinal temperature difference $\Delta T_x = T^+ - T^-$ (Fig. 2c). The thermal conductivity along the $x$ axis is given by $\kappa_{xx} = (Q / \Delta T_x) (L / wt)$, where $L$ is the separation (along $x$) between the two points at which $T^+$ and $T^-$ are measured, $w$ is the width of the sample (along $y$) and $t$ its thickness (along $z$). By applying a magnetic field $H$ along the $c$ axis of the crystal (along $z$), normal to the $CuO_2$ planes, one generates a transverse gradient $\Delta T_y$ (Fig. 2c). The thermal Hall conductivity is defined as $\kappa_{xy} = -\kappa_{yy} (\Delta T_y / \Delta T_x) (L / w)$, where $\kappa_{yy}$ is the longitudinal thermal conductivity along the $y$ axis. In this study, we take $\kappa_{yy} = \kappa_{xx}$. The thermal Hall conductivity $\kappa_{xy}$ of our samples was measured in magnetic fields up to $H = 18$ T. The measurement procedure is described in detail in ref. 22.

**Electrical measurements.** The longitudinal resistivity $\rho_{xx}$ and Hall resistivity $\rho_{xy}$ were measured in magnetic fields up to 16 T in a Quantum Design PPMS in Sherbrooke. (For Nd-LSCO $p = 0.20$, $\sigma_{xy}$ was measured at $H = 33$ T (ref. 14).) The measurements were performed using a conventional 6-point configuration with a current excitation of 2 mA, using the same contacts as for the thermal measurements (Fig. 2c). The electrical Hall conductivity $\sigma_{xy}$ is given by $\sigma_{xy} = \rho_{xy} / (\rho_{xx}^2 + \rho_{xy}^2)$.

FIELD DEPENDENCE OF THE THERMAL HALL CONDUCTIVITY

All of the data reported here were taken in a magnetic field (normal to the $CuO_2$ planes) large enough to fully suppress superconductivity, and thereby access the normal state of Nd-LSCO, Eu-LSCO, LSCO and Bi2201. Indeed, a field of 15 T is sufficient to do this in all samples presented here, down to at least 5 K. In the normal state, $\kappa_{xy}$ has an intrinsic field dependence. In Extended Data Fig. 4, we show how $\kappa_{xy}$ in LSCO $p = 0.06$, where $T_c = 5$ K, depends on magnetic field for $T > T_c$: the linear $H$ dependence of $\kappa_{xy}$ at high $T$ becomes sub-linear at low $T$.

THERMAL HALL CONDUCTIVITY IN YBCO

In YBCO at $p = 0.11$, there is huge negative $\kappa_{xy}$ signal in the field-induced normal state[22]. In this excellent metal, whose Fermi surface is reconstructed by CDW order into a small electron pocket of high mobility[13], the electrical Hall conductivity $\sigma_{xy}$ is equally huge. In fact, the WF law was found to hold, namely $\kappa_{xy} / T = L_0 \sigma_{xy}$ as $T \to 0$, within error bars of $\pm 15$ % (ref. 22). In other words, the negative $\kappa_{xy}$ signal in this case is due to the charge carriers (*i.e.* to electrons). However, because the ± 15 % uncertainty corresponds to ± 12 mW / K$^2$ m (in 27 T), it is impossible to know whether the $\kappa_{xy}$ signal in YBCO might also contain a contribution of – 2 to – 6 mW / K$^2$ m from neutral excitations    (*i.e.* – 1 to – 3 mW / K$^2$ m in 15 T; Fig. 1b).

THERMAL HALL SIGNAL FROM MAGNONS

In undoped La$_2$CuO$_4$, magnons have been well characterized by inelastic neutron scattering measurements[36]. There are two magnon branches, each with its own spin gap, of magnitude 26 K and 58 K, respectively. The thermal conductivity of magnons, $\kappa_{mag}$, is therefore thermally activated at $T$ < 26 K, so that $\kappa_{mag}$ decreases exponentially at low $T$. Hess *et al.* have estimated $\kappa_{mag}$ in La$_2$CuO$_4$ by taking the difference between in-plane and out-of-plane conductivities[37]. In Extended Data Fig. 5, we see that $\kappa_{mag}$ / $T$ decreases monotonically as $T \rightarrow$ 0 below 150 K.

By contrast, $\kappa_{xy}$ / $T$ in La$_2$CuO$_4$ increases monotonically with decreasing $T$, all the way down to $T$ ~ 5 K (Extended Data Fig. 5), a temperature 5 times smaller than the smallest gap, where there are no thermally excited magnons.

Moreover, when we move up in doping to $p$ = 0.06, where AF order is gone and LSCO is in a very different magnetic state (Fig. 1a), without well-defined magnons or a spin gap, $\kappa_{xy}(T)$ is essentially identical to that in La$_2$CuO$_4$ (Fig. 1b).

We conclude that magnons are not responsible for the large negative $\kappa_{xy}$ in cuprates.

THERMAL HALL SIGNAL FROM PHONONS

Phonons can produce a non-zero $\kappa_{xy}$ signal if they undergo skew scattering by spins[11,24]. Spin scattering of phonons can be detected through its impact on $\kappa_{xx}$. First, it reduces the magnitude of $\kappa_{xx}$ relative to its value without spin scattering. A good example of this is provided by the insulators Y$_2$Ti$_2$O$_7$ and Tb$_2$Ti$_2$O$_7$ . In non-magnetic Y$_2$Ti$_2$O$_7$ , $\kappa_{xx}(T)$ is large and typical of phonons in non-magnetic insulators (Extended Data Fig. 2a). In isostructural Tb$_2$Ti$_2$O$_7$, which has a large moment on the Tb ion, $\kappa_{xx}(T)$ is massively reduced (Extended Data Fig. 2a), as phonons undergo strong spin scattering. At $T$ = 15 K, $\kappa_{xx}$ is 15 times smaller in Tb$_2$Ti$_2$O$_7$.

A second and more direct signature of the spin scattering of phonons is a field dependence of $\kappa_{xx}$ . In Tb$_2$Ti$_2$O$_7$, a field of 8 T causes a 30% reduction in $\kappa_{xx}$ at $T$ = 15 K

(ref. 18; Fig. 4a, Extended Data Fig. 2b, Table 1). In the multiferroic material $(Fe,Zn)_2Mo_3O_8$, where the spin-phonon coupling is known to be very strong, a field of 9 T causes a 32% reduction in $\kappa_{xx}$ at $T = 30$ K (ref. 11; Fig. 4a, Table 1).

Let us now look for those two signatures in cuprates. First in Nd-LSCO, where the negative $\kappa_{xy}$ signal is absent at $p = 0.24$ and present at $p = 0.21$, with a magnitude ~ 10 times larger than in $Tb_2Ti_2O_7$. If this very large $\kappa_{xy}$ signal is due to phonons, then there must be some very strong spin scattering of phonons that appears below $p = 0.24$, which will show up as a massive decrease in $\kappa_{xx}$. In Extended Data Fig. 3, we see that there is no decrease of $\kappa_{xx}$ in going from $p = 0.24$ to $p = 0.21$, on the contrary, $\kappa_{xx}$ increases.

Secondly, let us look at the field dependence of $\kappa_{xx}$ in LSCO $p = 0.06$, where the negative $\kappa_{xy}$ signal is ~ 20 times larger than in $Tb_2Ti_2O_7$, at $T = 15$ K and $H = 8$ T (ref. 18; Extended Data Figs. 2b and 2d, Table 1). In LSCO, the change in $\kappa_{xx}$ induced by a field of 8 T at $T = 14$ K is no more than 1 % (Extended Data Figs. 1e and 2d), so ~ 20 times smaller than in $Tb_2Ti_2O_7$. In addition to being negligible in size, the $H$ dependence of $\kappa_{xx}$ in LSCO has the wrong $T$ dependence: $[\kappa_{xx}(15T) - \kappa_{xx}(1T)] / T$ drops below 30 K, whereas $\kappa_{xy} / T$ keeps growing monotonically as $T \to 0$ (Extended Data Fig. 1f).

We conclude that phonons are not responsible for the large negative $\kappa_{xy}$ in cuprates.

## EXTENDED DATA FIGURE CAPTIONS

**Extended Data Fig. 1 | Magnetic field dependence of $\kappa_{xx}$.**

Field dependence of $\kappa_{xx}$ in Eu-LSCO $p = 0.08$ (top panels) and LSCO $p = 0.06$ (bottom panels), displayed in three ways. **a), d)** $\kappa_{xx} / T$ vs $T$ at $H = 1$ T (blue) and $H = 15$ T (red). The difference between the two curves is very small, not visible by eye. **b), e)** Change in $\kappa_{xx}$ with field measured relative to its value at $H = 1$ T, $[\kappa_{xx}(H) - \kappa_{xx}(1\ T)]$ vs $H$, for various temperatures as indicated. **c), f)** Change in $\kappa_{xx}$ between 15 T and 1 T, plotted as $[\kappa_{xx}(H) - \kappa_{xx}(1\ T)] / T$ vs $T$ (blue, right axis), compared to $\kappa_{xy}(15\ T) / T$ vs $T$ (red, left axis). Note how at low $T$ the transverse response grows to be as large, if not larger, than the longitudinal response.

**Extended Data Fig. 2 | Comparing cuprates to pyrochlores.**

**a)** Thermal conductivity of two isostructural pyrocholore oxides, plotted as $\kappa_{xx} / T$ vs $T$ at $H = 0$, namely $Y_2Ti_2O_7$ (red) and $Tb_2Ti_2O_7$ (blue) (from (ref. 38). The presence of disordered magnetic moments in $Tb_2Ti_2O_7$ produces a strong scattering of phonons, seen as a massive suppression of $\kappa_{xx}$ (15-fold at $T = 15$ K). **b)** Field dependence of $\kappa_{xx}$, plotted as $\Delta\kappa_{xx}(H) / \kappa_{xx}(0)$ vs $H$, with $\Delta\kappa_{xx} = \kappa_{xx}(H) - \kappa_{xx}(0)$, at $T = 15$ K (blue; ref. 18). The strong effect of field (30% in 8 T) is a direct signature of the strong coupling between phonons and spins in $Tb_2Ti_2O_7$. Also shown is the transverse response in $Tb_2Ti_2O_7$ at $T = 15$ K, plotted as $\kappa_{xy} / T$ vs $H$ (red; ref. 18). Note that in $Y_2Ti_2O_7$, $\kappa_{xy} = 0$ (ref. 18). **c)** Thermal conductivity of two Nd-LSCO samples, on either side of $p^*$ (red, $p = 0.24$; blue, $p = 0.21$), plotted as $\kappa_{xx} / T$ vs $T$ at $H = 18$ T. We see that contrary to $Tb_2Ti_2O_7$ (panel a), the appearance of the negative $\kappa_{xy}$ signal in Nd-LSCO below $p^*$ is not accompanied by a large suppression of $\kappa_{xx}$, on the contrary (Extended Data Fig. 3). **d)** Same as in b), for LSCO $p = 0.06$, with the same $x$-axis and $y$-axis scales and data taken at (nearly) the same temperature. We see that the situation in LSCO is very different to that found in $Tb_2Ti_2O_7$ (panel b): instead of having a small $\kappa_{xy}$ and a large $\Delta\kappa_{xx}$ (panel b), we now have a large $\kappa_{xy}$ and a small $\Delta\kappa_{xx}$. Quantitatively, $\kappa_{xy} / \Delta\kappa_{xx} \sim 1$ in LSCO and $\sim 0.01$ in $Tb_2Ti_2O_7$, at $T = 15$ K and $H = 8$ T (Table 1).

**Extended Data Fig. 3 | Change in phonon $\kappa_{xx}$ across $p^*$ in Nd-LSCO.**

**a)** Thermal conductivity of Nd-LSCO at four different dopings, above $p^*$ ($p = 0.24$) and below $p^*$ ($p = 0.20, 0.21, 0.22$), plotted as $\kappa_{xx} / T$ vs $T$, at $H = 18$ T. We see that $\kappa_{xx}$ increases below $p^*$. **b)** Same as in panel a), for Nd-LSCO $p = 0.21$ (blue; $H = 18$ T) and LSCO $p = 0.06$ (green, $H = 16$ T). We see that $\kappa_{xx}$ continues to increase as we lower $p$ further. This shows that phonons conduct better at lower $p$. A natural explanation is that they are less scattered by charge carriers as the material becomes less metallic. **c)** Same data as in panel a), for Nd-LSCO $p = 0.21$ (blue) and $p = 0.24$ (red), compared to the electrical conductivity of those same samples, plotted as $L_0 / \rho$ vs $T$ (lines; measured at $H = 33$ T (ref. 14)). The latter curves are a reasonable estimate of the electronic thermal conductivity $\kappa_{xx}^{el}$, exact at $T \to 0$ (since the WF law is satisfied[39]), as seen in Fig. 2a. **d)** Estimate of the phonon conductivity, defined as $\kappa_{xx}^{ph} = \kappa_{xx} - L_0 T / \rho$, plotted as $\kappa_{xx}^{ph} / T$ vs $T$ (using data in panel c). We see that $\kappa_{xx}^{ph}(T)$ increases upon crossing below $p^*$, most probably because electron-phonon scattering is weakened by the loss of carrier density. There is no evidence that the phonons suddenly suffer from the onset of strong spin scattering below $p^*$ (which would cause $\kappa_{xx}^{ph}(T)$ to drop below $p^*$), such as would be required to explain the appearance of the negative $\kappa_{xy}$ signal below $p^*$ (Fig. 3) as being due to phonon transport.

**Extended Data Fig. 4 | Magnetic field dependence of $\kappa_{xy}$ in LSCO.**

**a)** Field dependence of the thermal Hall conductivity of LSCO at $p = 0.06$, plotted as $\kappa_{xy}$ vs $H$ at various temperatures, as indicated. The dependence of $\kappa_{xy}$ on $H$ is linear at high $T$ and it becomes sublinear at lower $T$. **b)** Deviation from linearity displayed by plotting $\kappa_{xy} / (T H)$ vs $T$ at four different fields, as indicated.

**Extended Data Fig. 5 | Magnon thermal conductivity in $La_2CuO_4$.**

**a)** Thermal conductivity of magnons in $La_2CuO_4$, plotted as $\kappa_{mag} / T$ vs $T$ (blue, right axis; ref. 37). The solid line is a fit to the data using the standard calculation for two magnon branches in 2D, with gaps as measured by neutron inelastic scattering[36], namely $\Delta_1 = 26$ K and $\Delta_2 = 58$ K. Below $T \sim 5$ K, thermally-excited magnons are exponentially rare and $\kappa_{mag} / T \sim 0$. In sharp contrast, the thermal Hall conductivity of $La_2CuO_4$, $| \kappa_{xy} / T |$ (red, left axis; Fig. 1b), is largest at $T \sim 5$ K. This comparison shows that the $\kappa_{xy}$ signal in $La_2CuO_4$ cannot come from magnon transport.

Figure 1

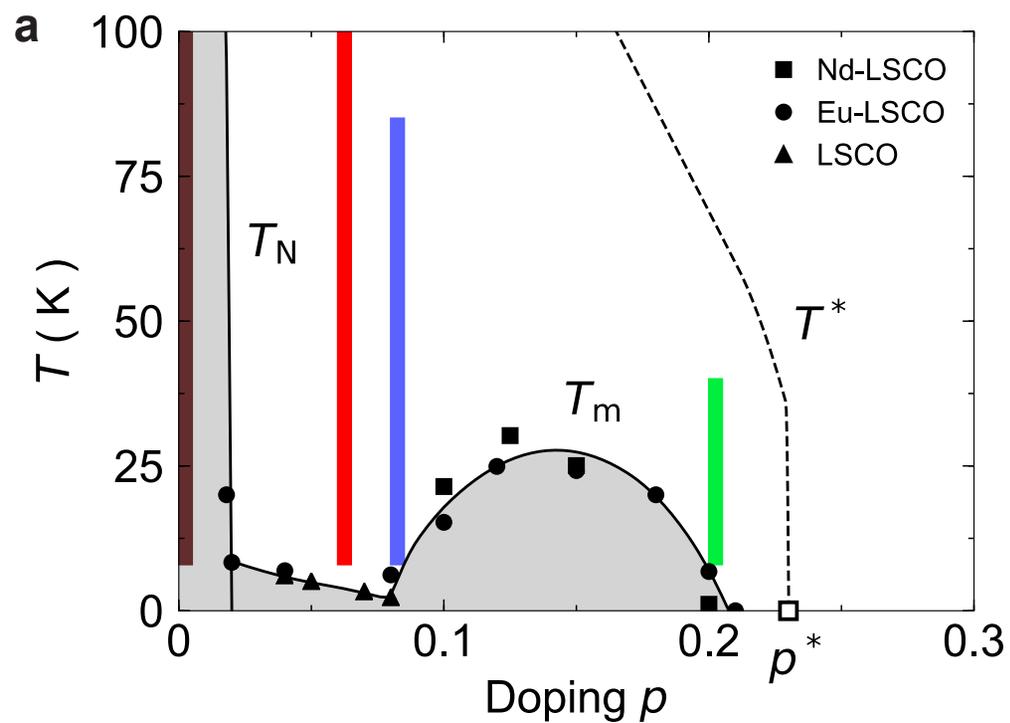
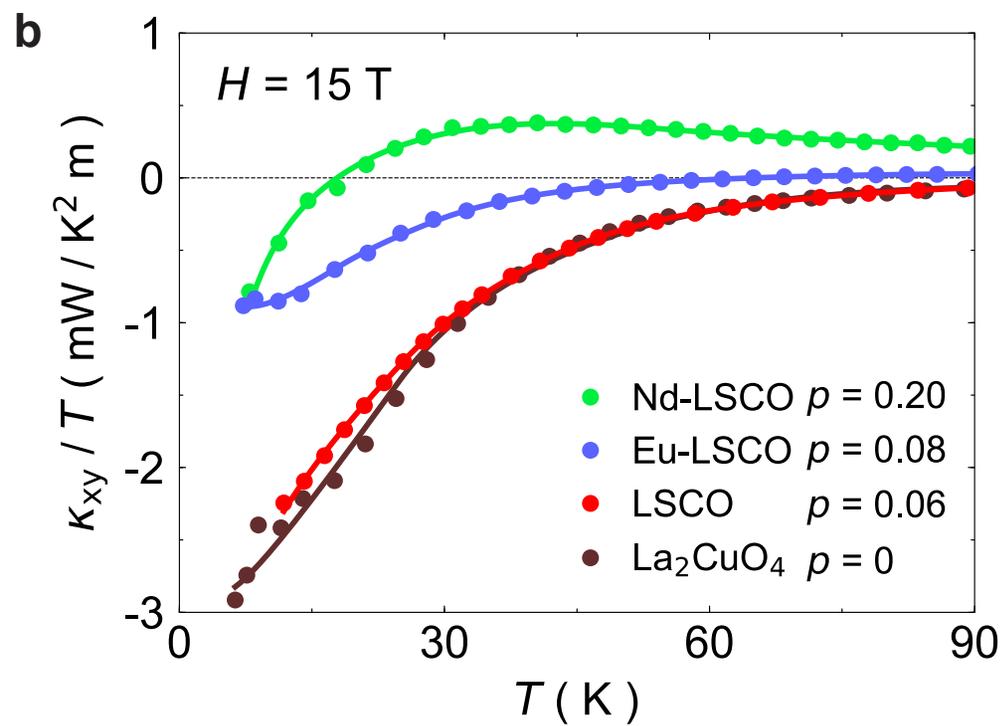

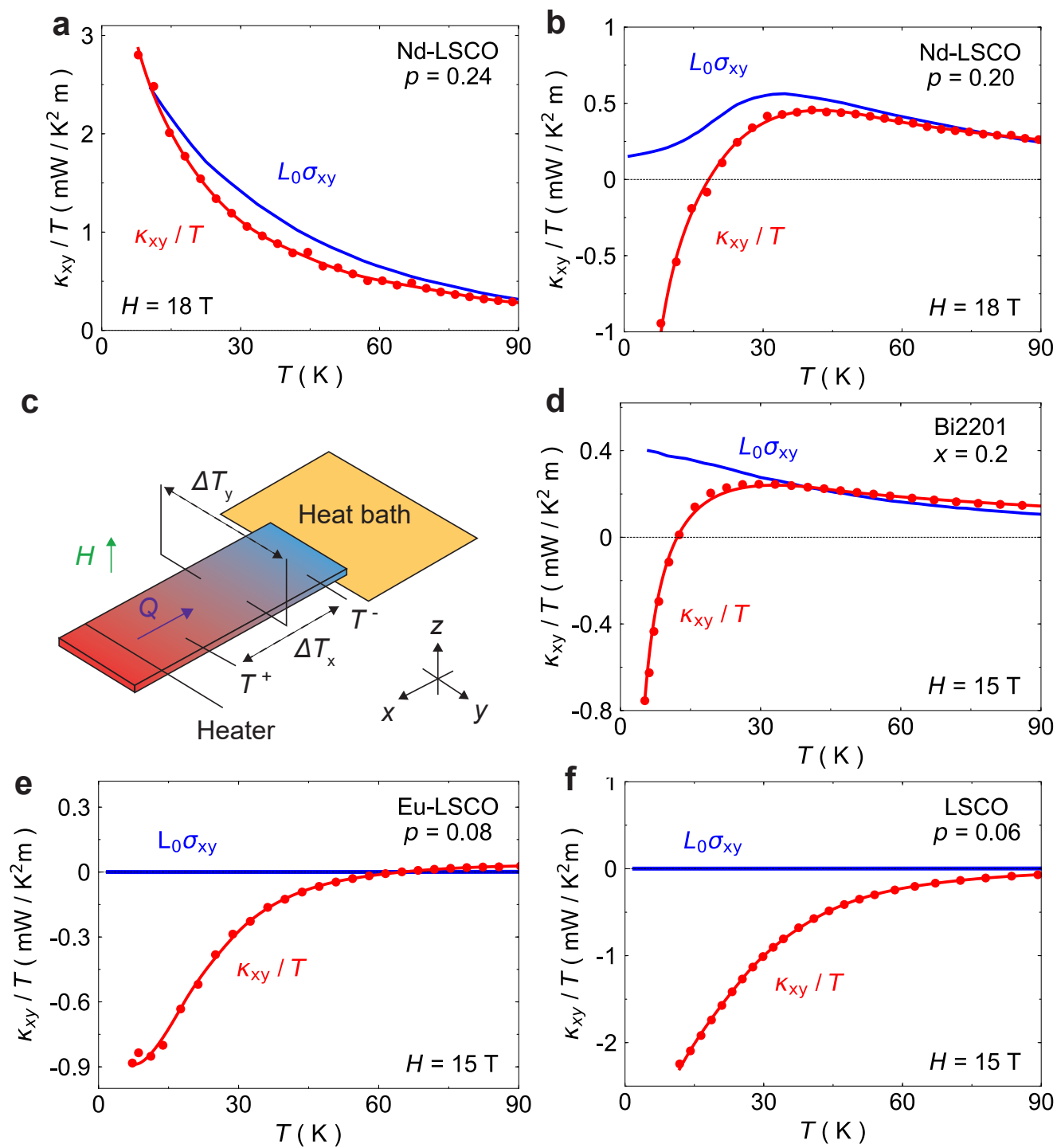

Figure 3

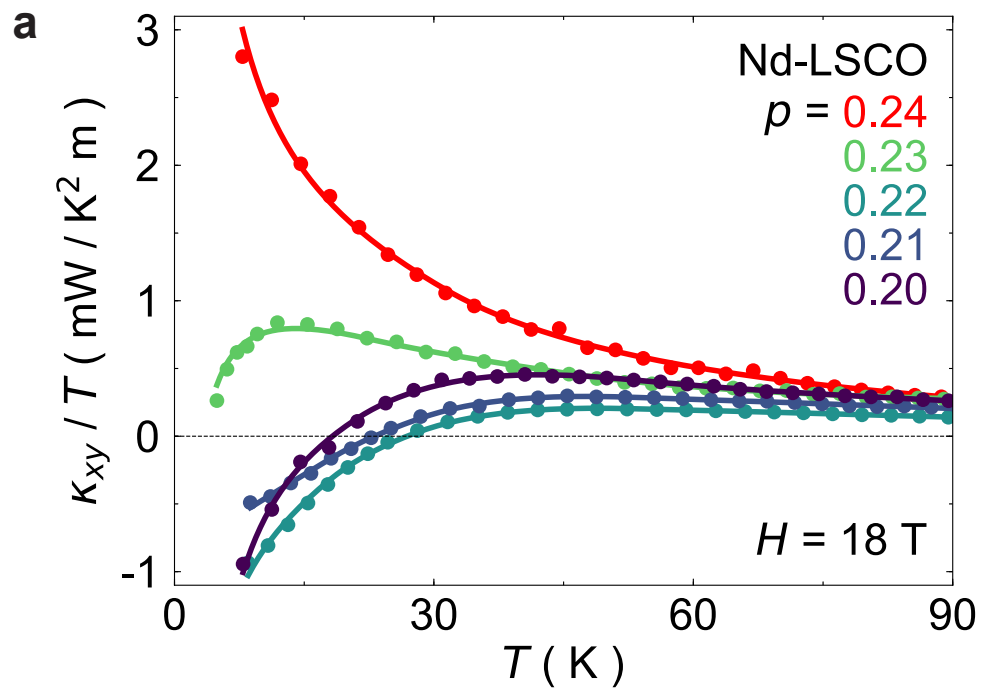
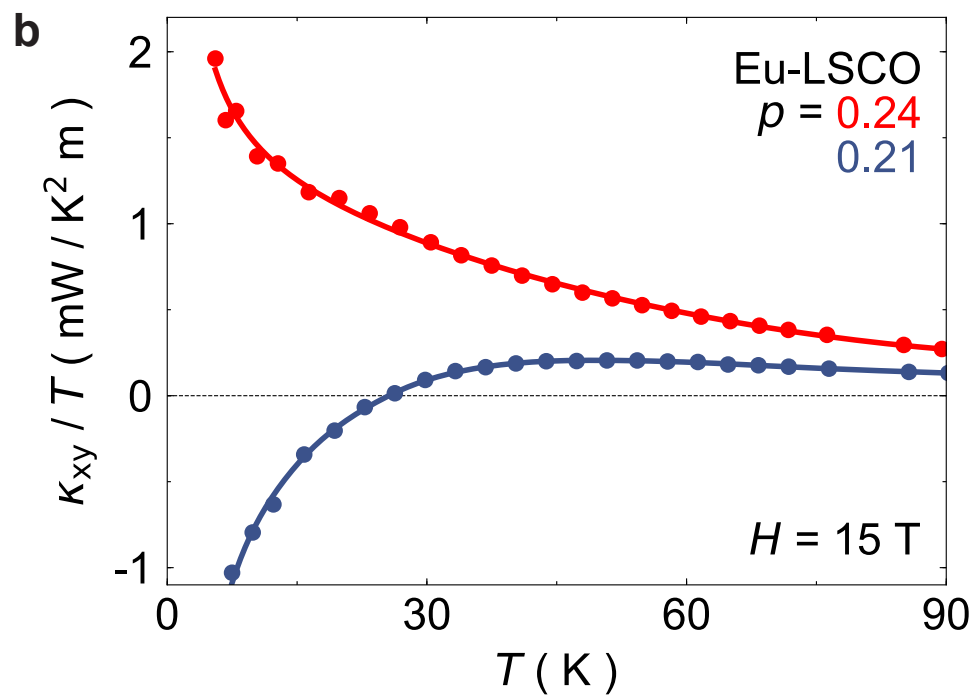

Figure 4

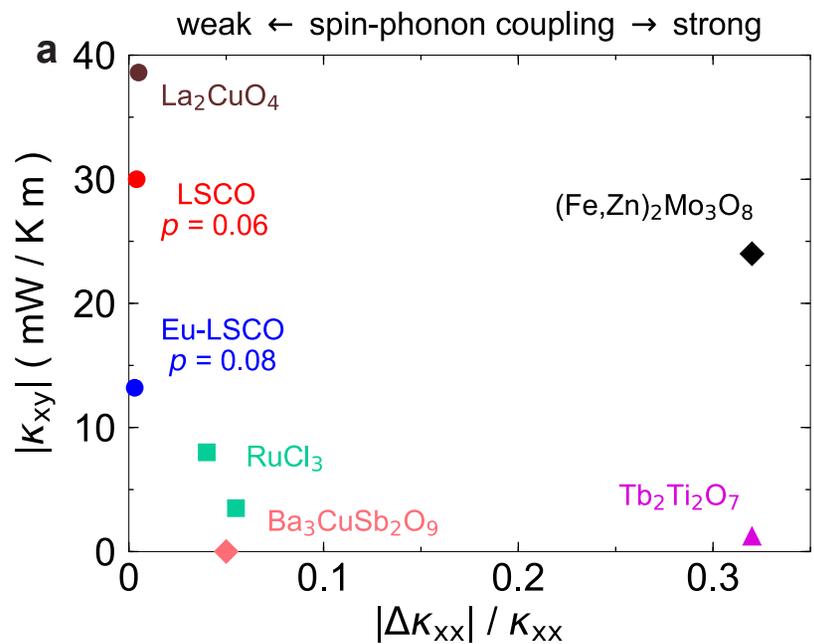
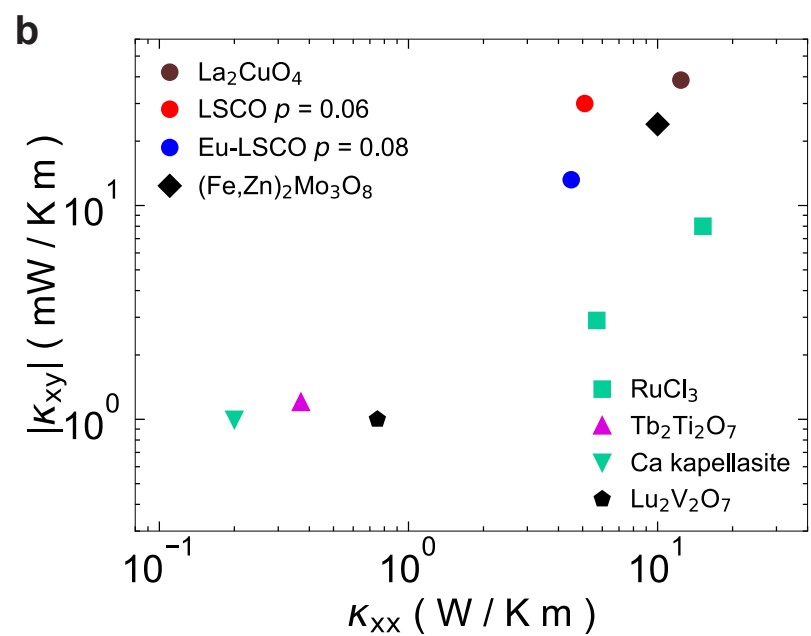
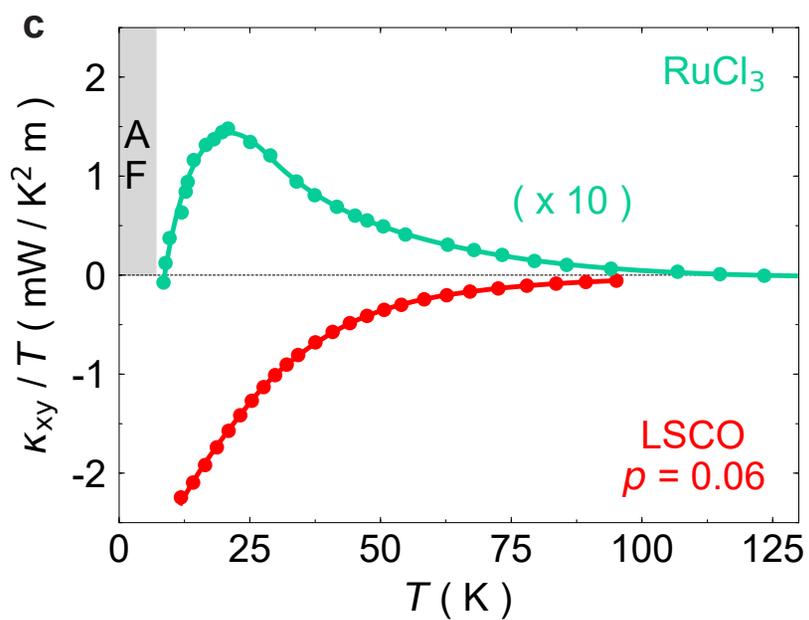
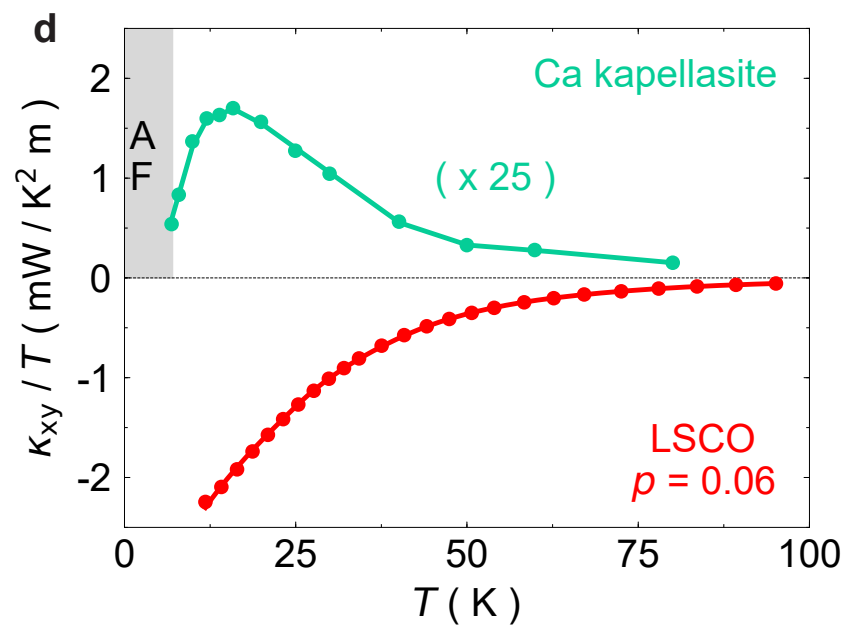

# Extended Data Figure 1

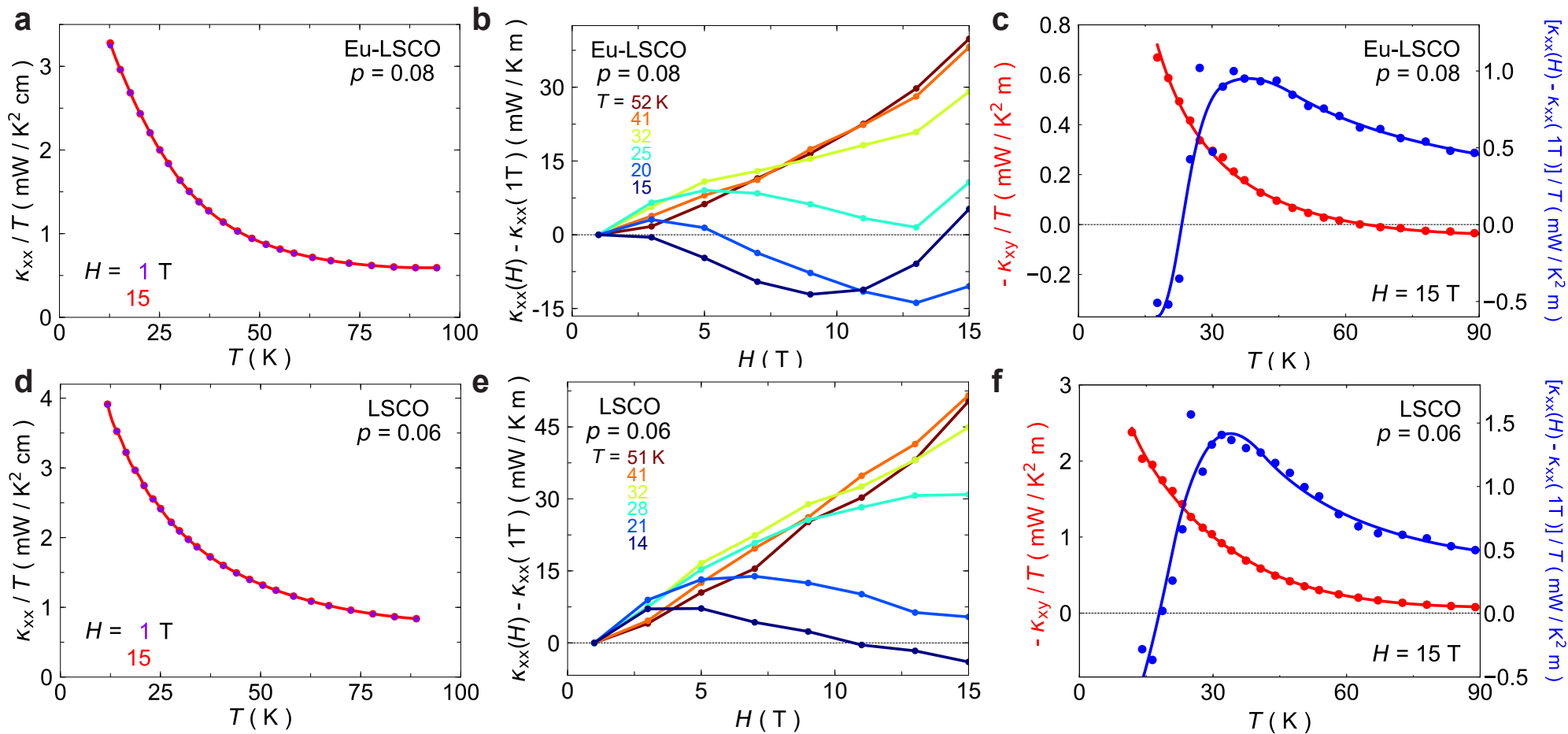

Extended Data Figure 2

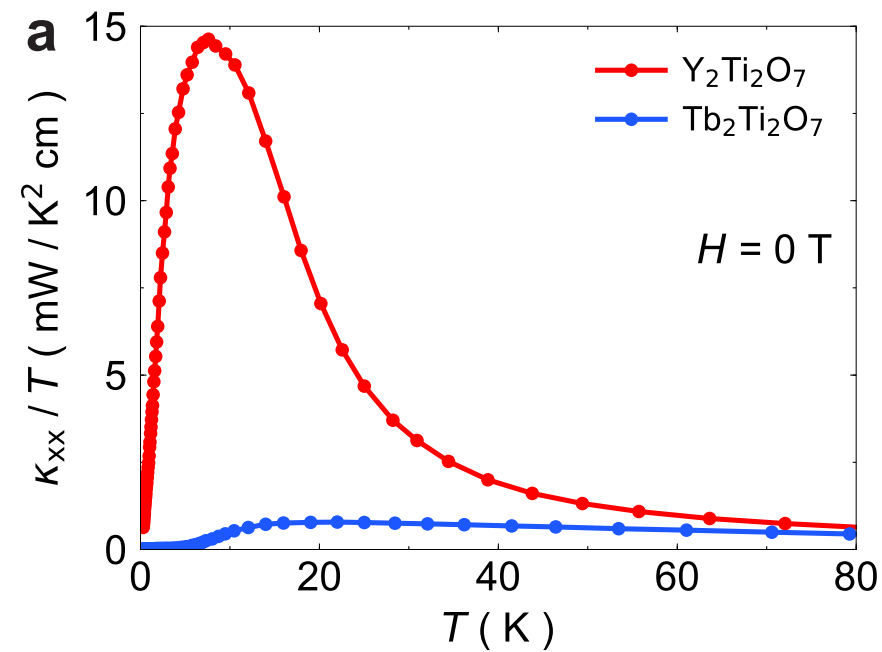
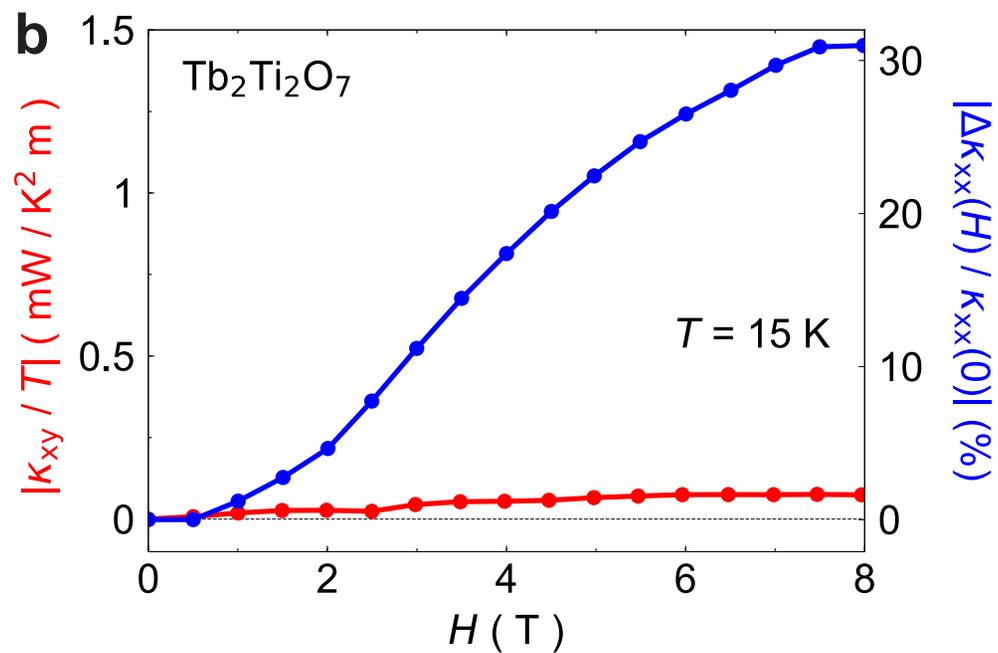
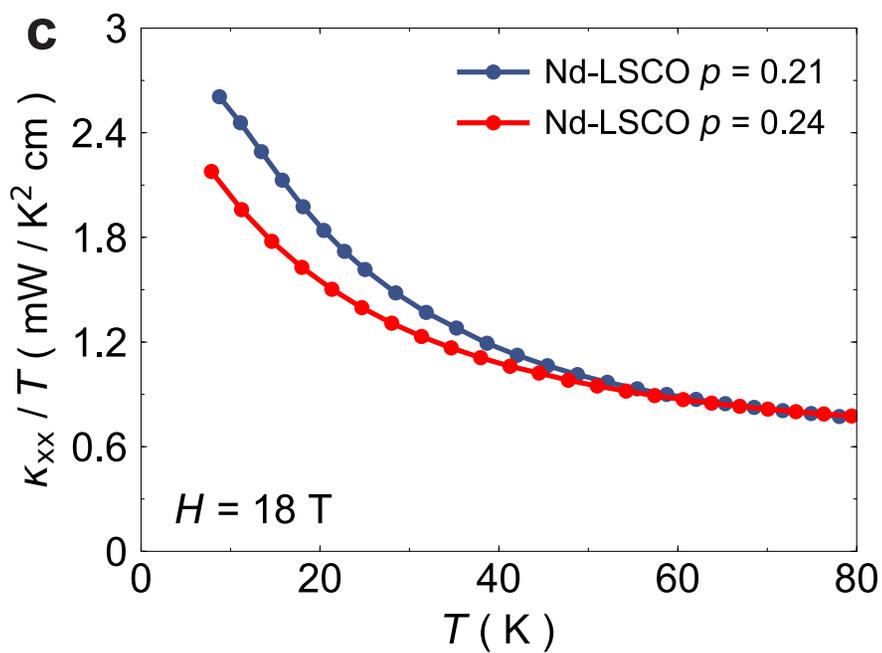
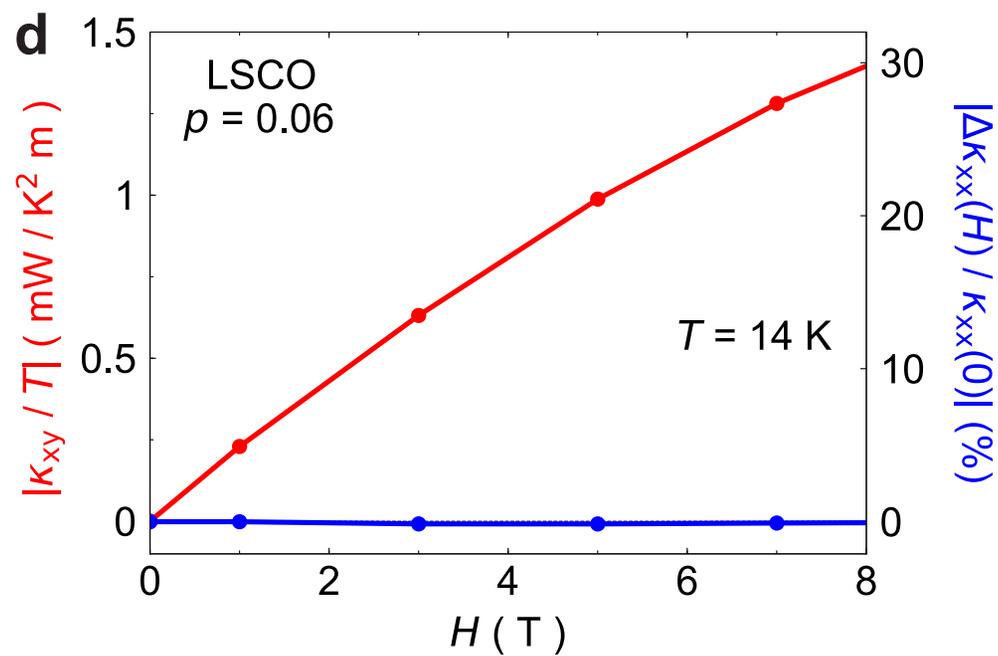

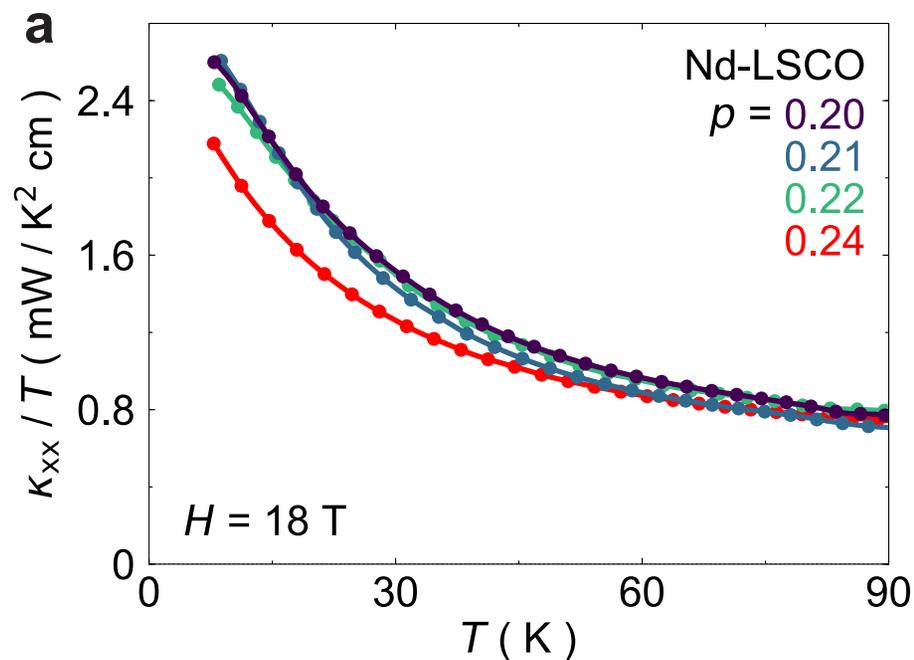
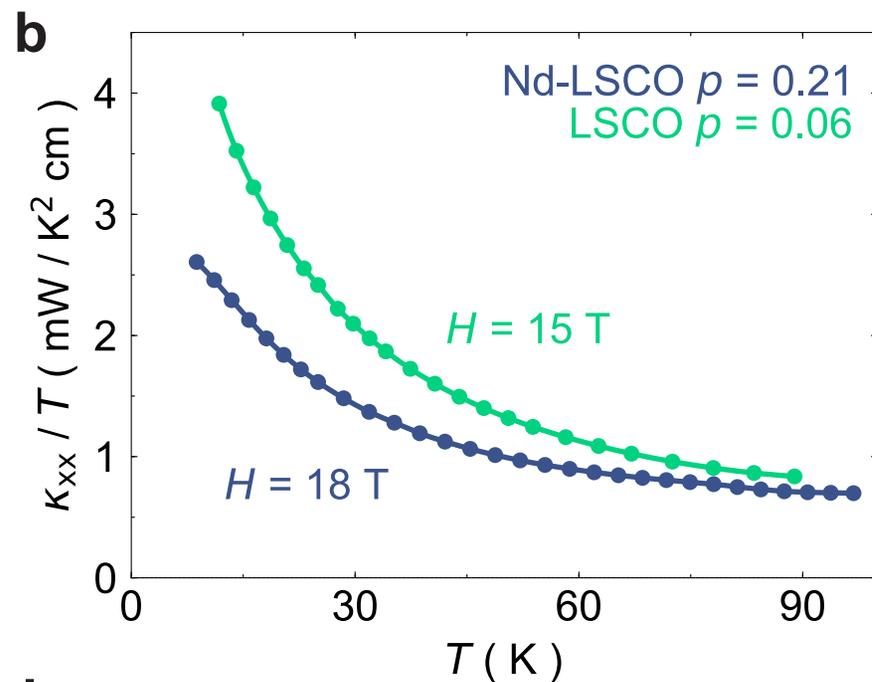
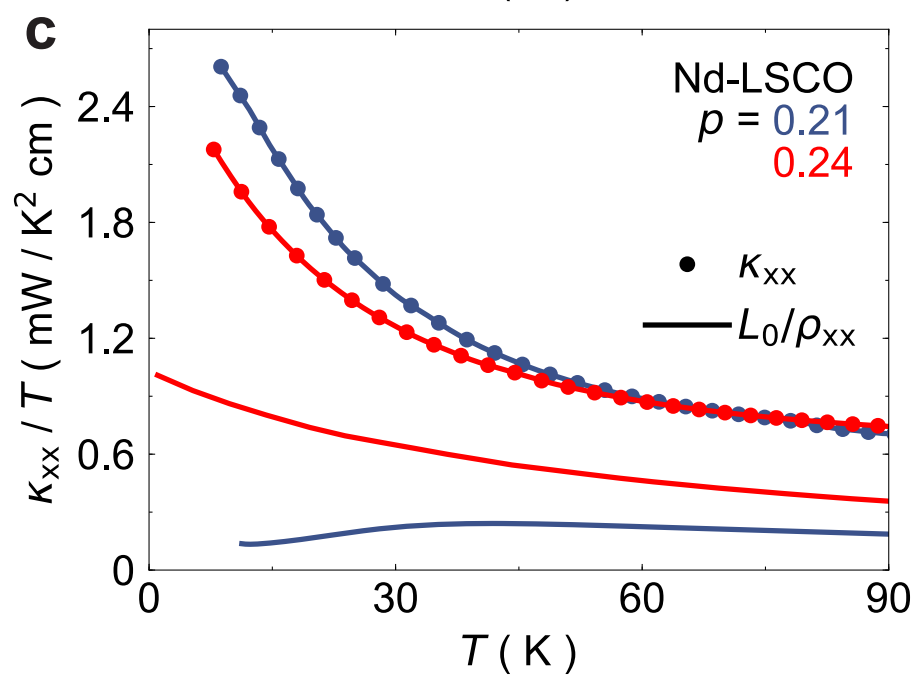
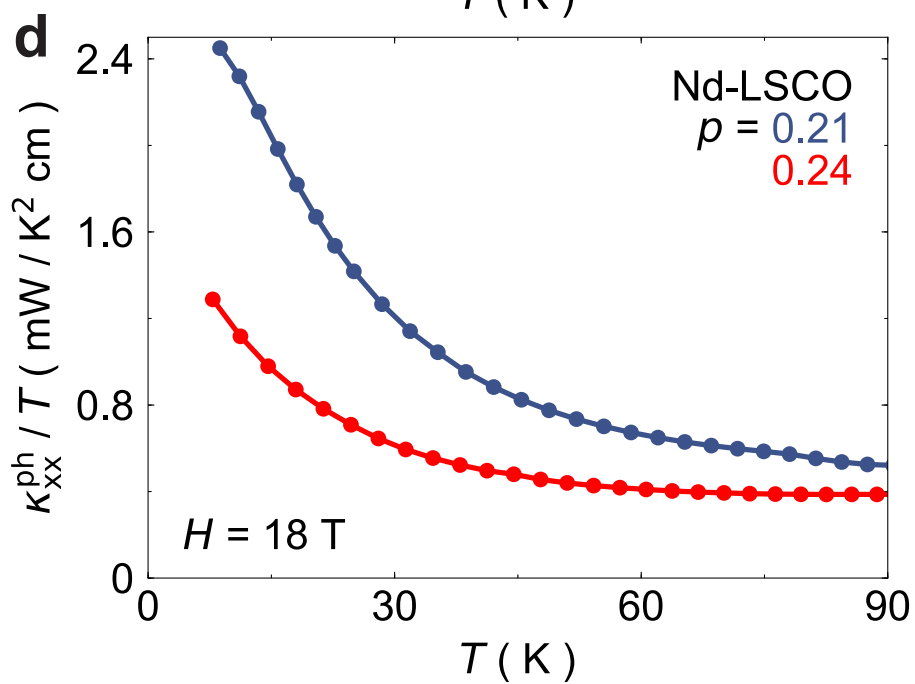

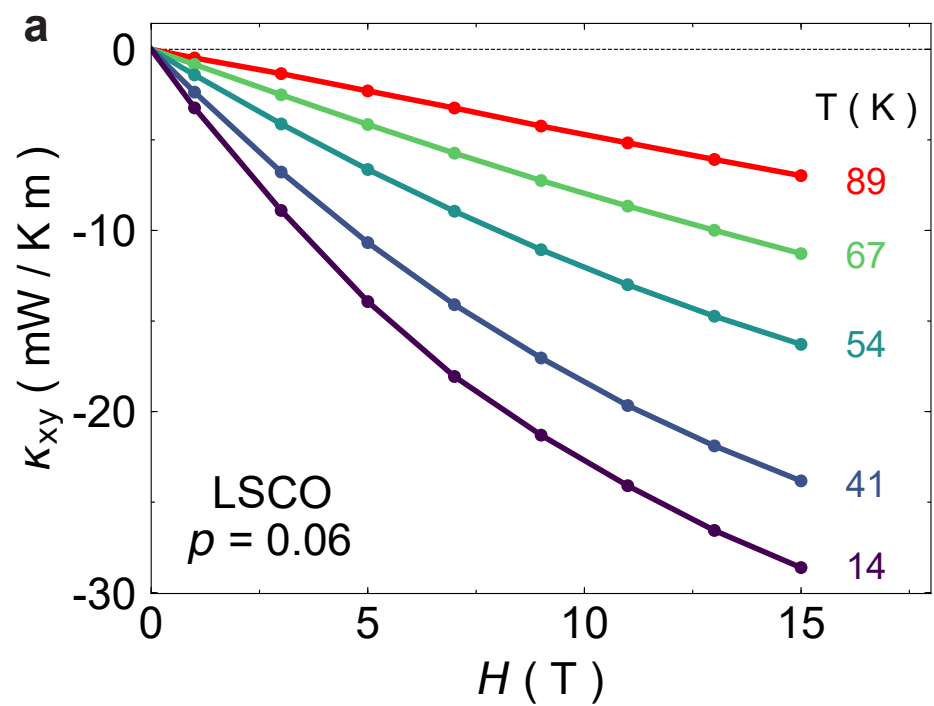
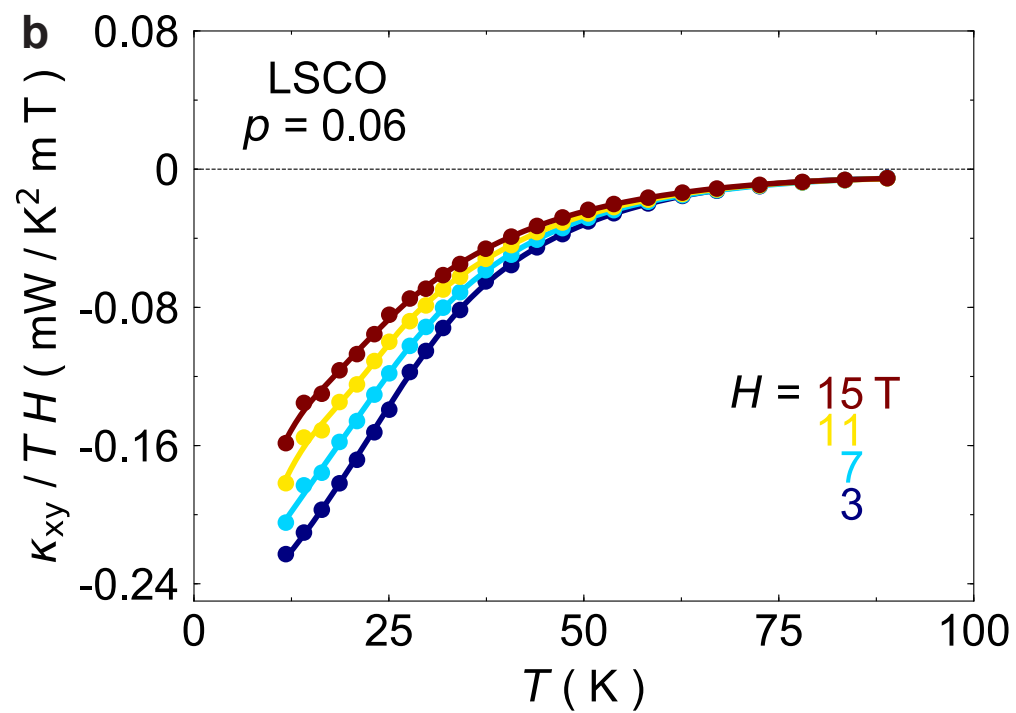

Extended Data Figure 5

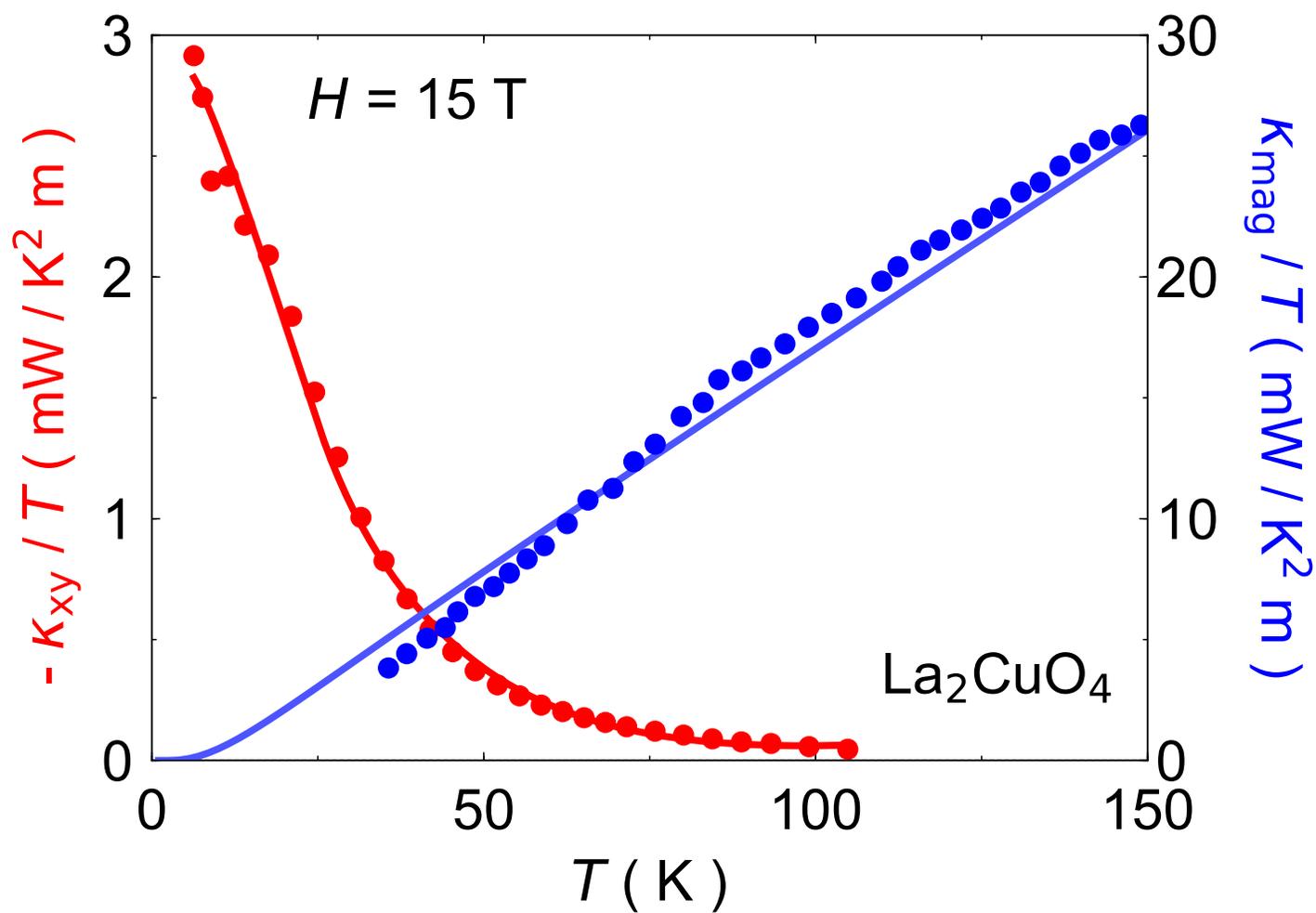